# First results of the two-phase argon avalanche detector performance with CsI photocathode

A. Bondar, A. Buzulutskov*, A. Grebenuk, D. Pavlyuchenko, R. Snopkov, Y. Tikhonov

*Budker Institute of Nuclear Physics, 630090 Novosibirsk, Russia*

**Abstract**

The performance of a two-phase Ar avalanche detector with CsI photocathode was studied, with regard to potential application in coherent neutrino-nucleus scattering and dark matter search experiments. The detector comprised a 1 cm thick liquid Ar layer and a triple-GEM multiplier operated in the saturated vapor above the liquid phase; the CsI photocathode was deposited on the first GEM. Successful detection of both primary scintillation and ionization signals, produced by beta-particles in liquid Ar, has for the first time been demonstrated in the two-phase avalanche mode.

*PACS:* 29.40.Cs; 34.80.My

*Keywords:* Two-phase avalanche detectors; Gas electron multipliers; Liquid argon; CsI photocathodes; Primary scintillations; Dark matter search

## 1. Introduction

Cryogenic two-phase detectors in general and specifically two-phase Ar detectors are currently favoured for application in low-background experiments, such as those for coherent neutrino-nucleus scattering [1], dark matter search [2,3] and medical imaging. To provide efficient background rejection in such experiments, both scintillation and ionization signals produced in the liquid by a particle need to be detected. It was shown that the detection of ionization signal can be achieved in Two-Phase Avalanche Detectors using Gas Electron Multipliers (GEMs) [4], operated in saturated vapour above the liquid phase [5]. This is provided by the unique ability of multi-GEM structures to operate in all noble gases at high gains at room [6] and cryogenic temperatures, including in the two-phase mode in Ar, Kr and Xe [5,7] (see also a review [8]).

It has also been recently demonstrated that a two-phase Ar avalanche detector with a triple-GEM multiplier can be operated at rather high gains, exceeding $10^4$ [7], including in a single electron counting mode [9]. These results are encouraging for the detection of scintillation signals in two-phase detectors, in particular using the GEM multiplier coupled to a photocathode. In this paper, the performance of a two-phase Ar avalanche detector with CsI photocathode is studied for the first time. We show that it is able to effectively detect scintillations produced by β-particles in liquid Ar.

## 2. Results

Three GEM foils, of an active area of 28×28 mm² each, were mounted in a cryogenic chamber of volume 2.5 l (Fig. 1). The detector was operated in Ar in the two-phase mode close to the triple point, at a temperature of 84 K and a vapor pressure of 0.70 atm. At this point, the liquid layer thickness was 10 mm. The electron drift path in liquid Ar was found to be substantially larger than 1 cm at a drift field of 1.5 kV/cm.

One more GEM foil was placed in the middle of the liquid layer; it was acted as an almost opaque cathode, with an optical transparency of 10%. The distance between the cathode and the first active GEM (GEM1 in Fig. 1) was 7 mm, between the GEMs - 2 mm and between the liquid surface and the first GEM - 2 mm. The triple-GEM multiplier was coupled to a reflective CsI photocathode following the idea described in [10,11]: the first GEM (gold-plated) was coated with a 2 μm thick CsI film using vacuum evaporation technique. The quantum efficiency of the CsI photocathode before installation into the chamber was measured to be about 5% at 185 nm. Accordingly, in the emission region of scintillations in Ar it is estimated to be of the same order, i.e. of 5%.

---

* Corresponding author. Tel.: +7-383-3394833; fax: +7-383-3307163.
  *E-mail address*: buzulu@inp.nsk.su

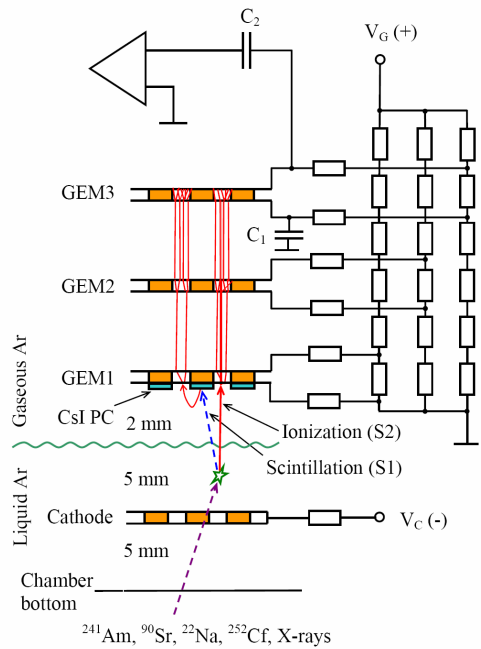

Fig. 1. Schematic view of the experimental setup to study the performance of the two-phase Ar avalanche detector with CsI photocathode (not to scale).

The cathode and GEM electrodes were biased through a resistive high-voltage divider placed outside the cryostat. The triple-GEM multiplier was operated in a "3GEM" mode: the anode signal was read out from the last electrode of the third GEM using a charge-sensitive preamplifier followed by a research amplifier (ORTEC 570) with a shaping time of either 0.5 or 10 μs and overall sensitivity of 11 V/pC in the latter case. The detector gain was measured using pulsed X-rays and the procedure described elsewhere [7]: it could exceed $10^4$.

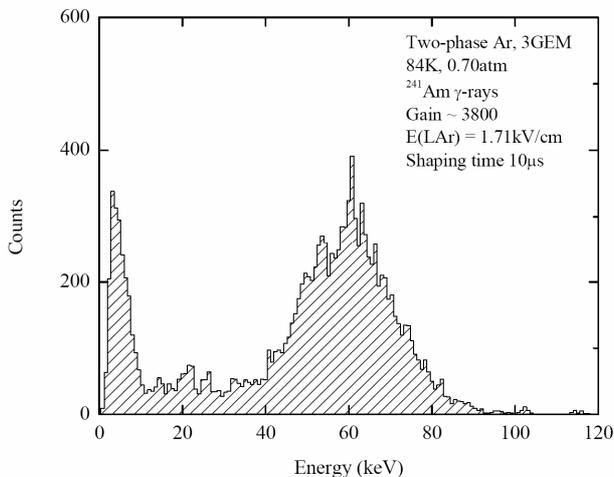

Fig. 2. Energy spectrum of $^{241}$Am γ-rays in the two-phase Ar avalanche detector at a gain of 3800, drift field E(LAr)=1.71 kV/cm and amplifier shaping time of 10 μs.

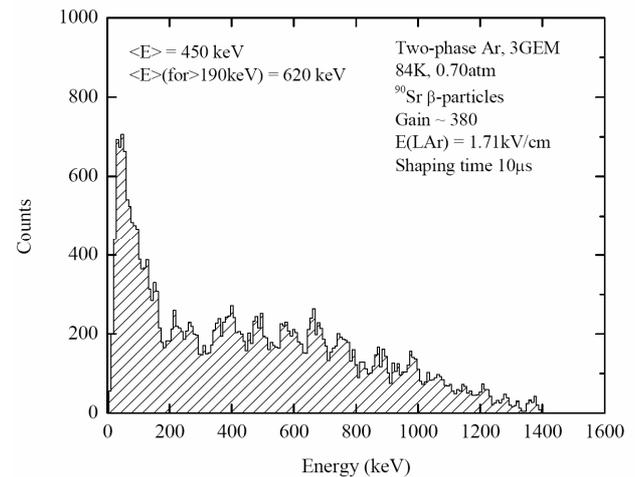

Fig. 3. Energy spectrum of $^{90}$Sr β-particles in the two-phase Ar avalanche detector at a gain of 380, drift field E(LAr)=1.71 kV/cm and amplifier shaping time of 10 μs.

The signal waveform analysis was carried out using TDS5032B digital oscilloscope: up to 5000 waveforms per measurement run could be stored in oscilloscope memory for offline processing. Other details of the experimental setup and procedures have been presented elsewhere [7,9].

The detector was irradiated by external sources, namely by $^{241}$Am γ-ray and $^{90}$Sr β-particle sources. Their energy spectra, as measured at a drift (extraction) field in liquid argon E(LAr)=1.71 kV/cm, are shown in Figs. 2 and 3. Here a 60 keV γ-ray peak from $^{241}$Am source was used to calibrate the energy scale. Since there was a 5 mm dead zone between the chamber bottom and the cathode, only a fraction of β-particle energy was deposited in the cathode gap. Therefore the average value of the high-energy component of the spectrum was about 600 keV only (Fig. 3). This should be compared to the value of 1 MeV for undisturbed β-particle spectrum of $^{90}$Sr. Note that β-particles are fully absorbed in 1 cm thick liquid Ar layer, i.e. cannot reach the gas gap and produce there ionization.

The idea of detecting both scintillation and ionization signals in the two-phase avalanche detector with CsI photocathode was suggested in [7,8]; it is illustrated in Fig. 1. The scintillation-induced photoelectrons released at the CsI photocathode are collected into the GEM holes and then multiplied, producing a so-called "S1" signal. The ionization-induced electrons are detected after some time, needed for drifting in the liquid and gas gaps and for emission through a liquid-gas interface; they produce the "S2" signal, delayed with respect to S1.

This concept was realized in the current work: we managed to observe both scintillation (S1) and ionization (S2) signals at a lower drift field, of 0.25 kV/cm, and smaller amplifier shaping time, of 0.5 μs. Such conditions were necessary to have enough time delay between S1 and S2 signals; otherwise they would overlap. Fig. 4 shows typical anode signals of a "S1+S2" type induced by β-particles in the two-phase Ar avalanche detector with CsI photocathode, at a gain of 5400. Here and in the following the trigger was defined by the S2 signal at a threshold of 100 mV. Fig. 5 shows a S1+S2 signal averaged over 111 events: the scintillation and ionization signals are distinctly seen and well separated.

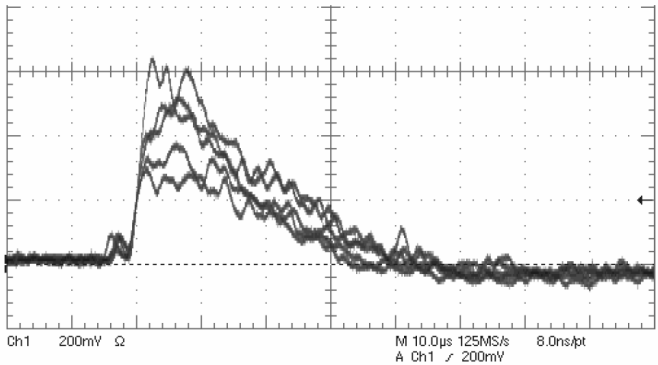

Fig. 4. Typical anode signals induced by $^{90}$Sr β-particles in the two-phase Ar avalanche detector with CsI photocathode at a gain of 5400, drift field E(LAr)=0.25 kV/cm and amplifier shaping time of 0.5 μs. The scintillation signals (S1), prior to the ionization signals (S2), are seen.

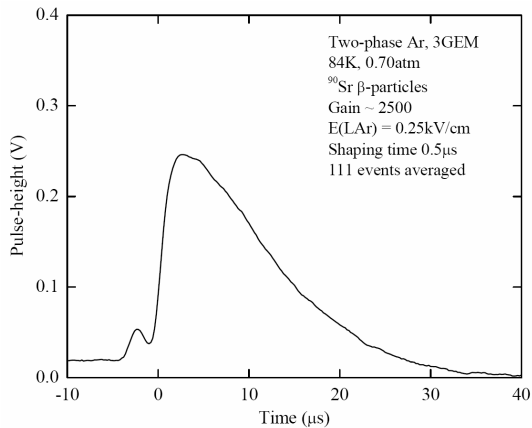

Fig. 5. Anode signal, averaged over 111 events of a S1+S2 type, induced by $^{90}$Sr β-particles in the two-phase Ar avalanche detector with CsI photocathode at a gain of 2500, drift field E(LAr)=0.25 kV/cm and amplifier shaping time of 0.5 μs. The scintillation (the first, S1) and ionization (the second, S2) signals are distinctly seen.

It should be noted that while the S1 signal was relatively fast, the S2 signal turned out to be rather slow: its width was as large as 30 μs on average. This is explained by the fact that at such a low extraction field (which in our case is equal to the drift field), the electron emission from liquid Ar is rather slow and is described solely by slow component [3]. Moreover, the electron emission efficiency here is substantially reduced, by more than an order of magnitude as for example compared to that at a field of 1.7 kV/cm. Nevertheless, the S2 signal is still considerably larger than the S1 signal.

In offline data analysis, a simple algorithm was developed to recognize the S1 signal: a maximum (a peak), followed by a minimum, was looked for within a certain time interval, prior to the S2 signal. Fig. 6 shows peak delay spectra of the S1 signal with respect to the S2 trigger at drift fields of 0.25 and 0.61 kV/cm. One can see that the time delay between S1 and S2 depends on the drift field and is larger for lower fields. At a drift field of 0.25 kV/cm its most probable value is 2.4 μs, which corresponds to electron drifting through the liquid and gas gaps. This confirms the statement that the S1 signal is indeed induced by primary scintillations.

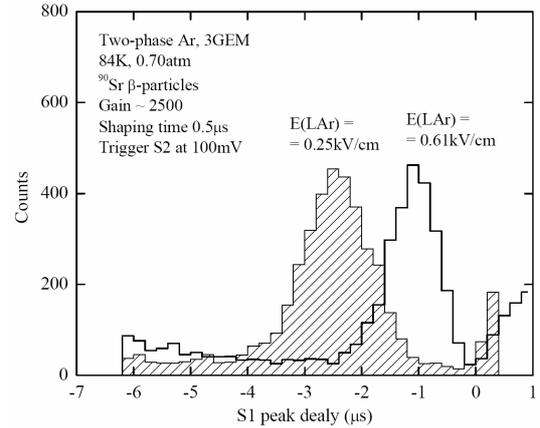

Fig. 6. Peak delay spectra of S1 signal with respect to S2 trigger for different drift fields in liquid Ar. The signals are induced by $^{90}$Sr β-particles in the two-phase Ar avalanche detector with CsI photocathode at a gain of 2500 and amplifier shaping time of 0.5 μs.

The amplitude of the S1 or S2 signal is obviously proportional to the time integral of the signal; it was determined by calculating the area under the appropriate part of the waveform. Fig. 7 shows the distribution of events in the plane S2 versus S1 amplitudes, at a gain of 2500. One can see that most events are of "S1+S2" type where both S1 and S2 signals are observed. Note that they are correlated (proportional) to each other. On the other hand, some events have a S2 signal alone, presumably due to a reduced geometrical acceptance for scintillation detection in some cases, for example in those where ionization is produced in the corner of the active area. Here the "S1" signal is just a noise. A negligible amount of events are due to S2 treated as S1: in Fig. 6 these are contained in the tail of events with a positive time delay.

Fig. 8 shows amplitude spectra of S1 and S2 signals. The S2 spectrum has a single peak corresponding to high-energy component of the β-particle spectrum in Fig. 3; low-energy component was disregarded since it was under the trigger threshold. The S1 spectrum has also a peak, obviously corresponding to S2 peak. In addition it has a tail at lower amplitudes, corresponding to electronic noise. This tail reflects the events of "S2 alone" type in Fig. 7.

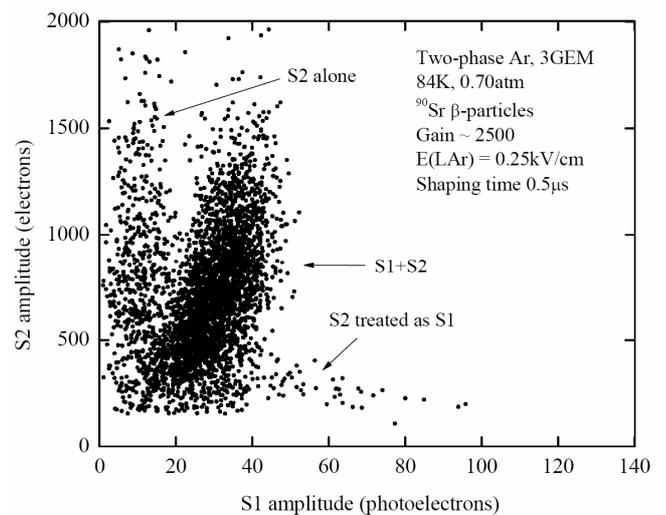

Fig. 7. Distribution of events in the plane of S2 versus S1 amplitudes. The signals are induced by $^{90}$Sr β-particles in the two-phase Ar avalanche detector with CsI photocathode at a gain of 2500, drift field E(LAr)=0.25 kV/cm and amplifier shaping time of 0.5 μs.

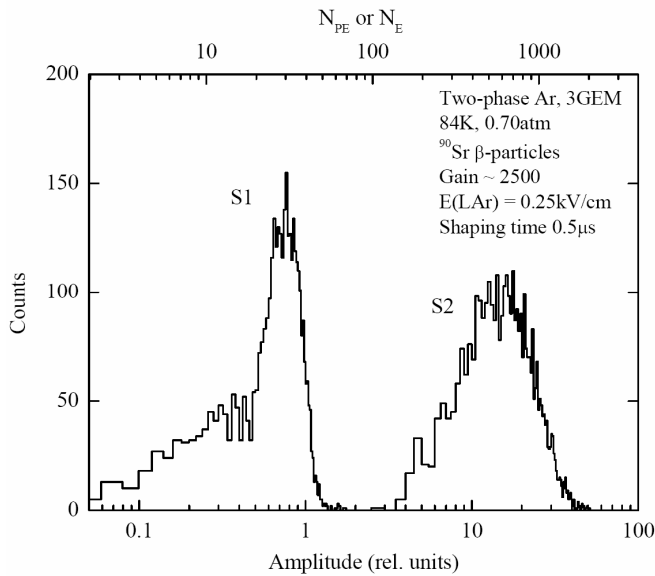

Fig. 8. Amplitude spectra of S1 and S2 signals from $^{90}$Sr β-particles in the two-phase Ar avalanche detector with CsI photocathode at a gain of 2500, drift field E(LAr)=0.25 kV/cm and amplifier shaping time of 0.5 μs. The top scale is expressed in initial charge prior multiplication, i.e. in photoelectrons for S1 and electrons for S2.

In Figs. 7 and 8 (top scale), the S1 and S2 scales are expressed in the initial charge prior to multiplication, i.e. in photoelectrons and electrons respectively. The appropriate scale calibration was carried out using the gain value and amplifier calibration at a given shaping time. It should be noted however, that the accuracy of this calibration procedure was not too high: of a factor of 1.5. More details will be presented elsewhere in a more elaborated paper [12].

As one can see from Fig. 8, the number of photoelectrons in the S1 peak is about $N_{PE}$=30. This corresponds to a deposited energy of about 600 keV. Hence, the photon detection efficiency of the CsI/triple-GEM assembly can be estimated: $\varepsilon=N_{PE}/N_{PH}$, where $N_{PH}$ is the number of scintillation photons emitted. Accounting for the scintillation light yield in liquid Ar, of 40 photons per one keV [13], the efficiency is estimated to be of the order of $\varepsilon \sim 10^{-3}$.

It should be noted that this estimation gives only a lower limit of the photon detection efficiency, in particular due to the fact that the scintillation light yield decreases with the drift field. Nevertheless, compared to the quantum efficiency of the CsI photocathode itself, the photon detection efficiency is substantially reduced, by a factor of 50. This factor includes the geometrical acceptance, the effect of photoelectron backscattering to CsI photocathode [8], which is rather strong in pure noble gases, and the photoelectron collection efficiency into the GEM holes [10].

## 3. Conclusions

The performance of a two-phase Ar avalanche detector with CsI photocathode was studied. The detector comprised a 1 cm thick liquid Ar layer, a triple-GEM multiplier operated in saturated vapor above the liquid phase and a CsI photocathode deposited on the first GEM. Successful detection of both primary scintillation and ionization signals, produced by beta-particles in liquid Ar, has for the first time been demonstrated in the two-phase avalanche mode. The amplitude of the scintillation signal is estimated to be about 30 photoelectrons per 600 keV of deposited energy.

The scintillation photon detection efficiency is estimated to be of the order of $10^{-3}$ and is presumably limited by the effects of photoelectron backscattering and collection. Therefore, the improved detector should be optimized in terms of photoelectron extraction and collection efficiencies.

The results obtained are relevant in the field of low-background detectors sensitive to nuclear recoils, such as those for coherent neutrino-nucleus scattering and dark matter search experiments. Further studies of this technique are in progress.

The research described in this publication was made possible in part by an INTAS Grant, award 04-78-6744.